\documentclass[aps,prb,showpacs,superscriptaddress,twocolumn]{revtex4}

\usepackage{graphicx,amssymb,amsfonts,amsmath}
\usepackage{natbib,hyperref}%to get nice hyperlinks in the references

\def \fig {{fig}}

\usepackage[latin1]{inputenc}
\usepackage{amsmath}
\usepackage{amsfonts}
\usepackage{amssymb}
\usepackage{graphicx}
\usepackage[usenames]{color}
\usepackage{epstopdf}
\newcommand{\be}{\begin{equation}}
\newcommand{\ee}{\end{equation}}

\def \be{\begin{equation}}
\def \ee{\end{equation}}
\def \ba{\begin{array}}
\def \ea{\end{array}}
\def \bea{\begin{eqnarray}}
\def \eea{\end{eqnarray}}
\def \nn{\nonumber}
\def \half{{1\over 2}}

\def \D{{\Delta}}

\def \w{{\omega}}

\def \av#1{{\langle#1\rangle}}

\def \beas{\begin{eqnarray*}}
\def \eeas{\end{eqnarray*}}

\def \half{{\frac{1}{2}}}

\newcounter{indice}

\def \bn{\begin{enumerate}}
\def \en{\end{enumerate}}

\def \bb{\begin{thebibliography}{99}}
\def \eb{\end{thebibliography}}

\begin{document}
\title{Noisy quantum phase transitions: an intuitive approach}

\author{Emanuele G. Dalla Torre}
\affiliation{Department of Physics, Harvard University, Cambridge MA
02138}
\author{Eugene Demler}
\affiliation{Department of Physics, Harvard University, Cambridge MA
02138}
\author{Thierry Giamarchi}
\affiliation{DPMC-MaNEP, University of Geneva, 24 Quai Ernest-Ansermet, 1211 Geneva,
Switzerland}
\author{Ehud Altman}
\affiliation{Department of Condensed Matter Physics, Weizmann Institute of Science,
Rehovot, 76100, Israel}


\begin{abstract}
Equilibrium thermal noise is known to destroy any quantum phase transition. What are the effects of non-equilibrium noise? In two recent papers we have considered the specific case of a resistively-shunted Josephson junction driven by $1/f$ charge noise. At equilibrium, this system undergoes a sharp quantum phase transition at a critical value of the shunt resistance. By applying a real-time renormalization group (RG) approach, we found that the noise has three main effects: It shifts the phase transition, renormalizes the resistance, and generates an effective temperature. In this paper  we explain how to understand these effects using simpler arguments, based on Kirchhoff laws and time-dependent perturbation theory. We also show how these effects modify physical observables and especially the current-voltage characteristic of the junction. In the appendix we describe two possible realizations of the model with ultracold atoms confined to one dimension.
\end{abstract}

\maketitle

\section{Introduction}

Recent years have seen increasing interest in non-equilibrium many-body quantum systems. Schematically, these systems can be divided in two categories: ``closed'', if isolated from the environment, and ``open'' if coupled to an external environment. In closed systems, a non-equilibrium situation can be generated by preparing the system in a given many-body quantum state and letting it evolve according to its (time-independent) Hamiltonian. The resulting dynamics depends of course on both the initial state and the Hamiltonian, and is therefore highly non-universal. To allow universal predictions, the initial state is often chosen as the ground state of some particular Hamiltonian. For instance, if this Hamiltonian is close to a quantum critical point, the dynamics of the system is expected to follow universal scaling laws\cite{Cardy,DeGrandi,Sengupta}.

In open quantum systems, one is often not interested in the dynamics of the system, but rather in the properties of a non-equilibrium steady state, arising due to the flow of energy between two (or more) baths. In many cases, one bath is modeled in terms of a classical force, such as a voltage source, or an optical pump, while the second bath is treated in a full quantum manner. Depending on the details of the system, the quantum bath can be either Markovian, as usually assumed in quantum optics, or non-Markovian, as more common in solid state devices. This difference gives rise to a different formalism needed: quantum master equation in the former case\cite{DiehlZoller,Prosen,Gritsev}, and non-local real-time actions in the latter\cite{MitraMillis,MitraMillis2}.

Here we focus on non-Markovian quantum baths and, in particular, zero-temperature Gaussian baths. This type of baths can be obtained by integrating-out an infinite set of harmonic oscillators, initially prepared in their ground state\cite{FeynmanVernon,CaldeiraLeggett}. In the absence of an external pump, this problem has been widely studied in the literature, especially in the context of quantum phase transitions\cite{Schmid,Chakravarty,FisherZwerger,LeggettReview,KaneFisher}. The canonical example is a quantum particle in a double-well or in a periodic potential, under the effects of an Ohmic dissipative bath. If the coupling to the bath is weak, the particle occupies a coherent superposition of all minima of the potential. When the coupling to the bath becomes strong enough, the particle localizes in one minimum of the potential, hence breaking the initial symmetry of the problem, through a universal quantum phase transition. In the case of a periodic potential, this transition corresponds the insulator-superconductor quantum phase transition of a single resistively-shunted Josephson junction. The effects of an external drive on this systems is the subject of this work.

In two recent papers\cite{us-nature,us-prb}, we studied the steady state of a resistively-shunted Josephson junction driven by a stochastic voltage source, corresponding to $1/f$ charge noise. Exploiting the scale invariance of the problem, we developed a novel analytic real-time renormalization group (RG) approach\cite{us-nature,Mitra-short,us-prb,Mitra-long}. This approach offers a  controlled way to describe the low-voltage properties of the junction. However, being expressed in terms of Keldysh path-integrals, it may appear highly non-transparent to the reader who is not familiar with this formalism. The goal of this paper is to study the same problem using simpler methods. Specifically, we will substantially relay on circuit theory, Kirchoff laws, and perturbation theory. Yet, we will reproduce all the main results obtained from the more involved RG calculations. 

This article is organized as follows. In Sec. \ref{sec:model} we will introduce the specific model considered here, a resistively-shunted Josephson junction driven by $1/f$ charge noise. The following three sections are devoted to the three main effects of the noise, namely: the shift of the transition (Sec. \ref{sec:first}), the renormalization of the resistance  (Sec. \ref{sec:second}), and  the generation of an effective temperature (Sec. \ref{sec:third}). In Sec. \ref{sec:IVcurve} we explain how these effects concur to determine the non linear current-voltage characteristic of the junction. Sec. \ref{sec:summary} concludes the article with a brief summary and open questions. In the  appendix we describe a possible realization of the model using ultracold atoms in one dimension.

\section{The model: a noise-driven superconducting junction}\label{sec:model}

\begin{figure}
\centering
\includegraphics{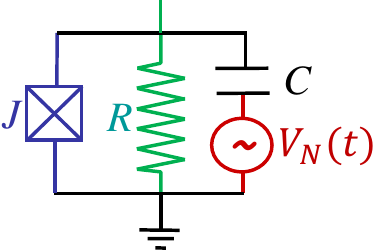}
\caption{The noise-driven resistively shunted Josephson junction: (a) Electric circuit of the junction. (b) Realizations with ultracold atoms (see Appendix)}
\label{fig:circuit}
\end{figure}

The object of the present study is the non-equilibrium device plotted in Fig. \ref{fig:circuit}. As explained in the introduction, the circuit consists of three main elements:

1) A ``pump''. A stochastic, time-dependent voltage source $V_{N}(t)$, capacitively coupled to the resistor. This voltage source models the so called ``charge noise'', due to time-dependent fluctuations of charges in the substrate \footnote{Apart from charge noise, superconducting junctions are affected also by ``current noise'' or ``flux noise'', due to time dependent fluctuations of the Josephson coupling. This type of noise is highly non-linear, hard to predict, and will be neglected in this work.}. 
Being coupled linearly to the circuit, this type of noise has two main advantages: it can be easily introduced from the outside in a controlled experiment, and it allows for an exact analytical treatment as explained below. In the following, we will consider in particular the case of $1/f$ noise spectrum $\av{V_N(\w)V_N(\w')} = F_0 / (2\pi|\w|) \delta(\w-\w')$, where $F_0$ measures the strength of the noise and has units of voltage-square. This parameter can be combined with the capacitance $C$ and the electric charge (of a Cooper pair) $2e$ to form a unitless parameter $F = F_0 C^2 / (2e)^2$. As we will see this parameter plays an important role in physics of the problem. %In typical circuits \cite{Zimmerli} $1/f$ noise is highly suppressed and this parameter is of order $10^{-5}$. We deduce that, in order to observe the effects described in this article, additional external noise should be introduced from the outside.

2) A ``dissipative bath''. A linear resistor, corresponding to Ohmic dissipation, which is assumed to be at equilibrium (and specifically at zero temperature). The coupling to the bath is measured by the unitless ratio $R/R_Q$, where $R_Q=h/(2e)^2$ is the quantum resistance. The regimes $R/R_Q<1$ and $R/R_Q>1$ are often termed respectively ``overdamped'' and ``underdamped''. The equilibrium quantum phase transition sits precisely at the boundary between these two regimes.

3) A ``quantum system''. A non-linear tunneling junction, obeying the Josephson relations $V(t)=(h/2e)\partial_t \phi(t)$ and $I(t)=J\sin(\phi)$, where $h$ is the Plank constant (in order to fulfill our didactic goal we shall not put $\hbar=1$ as often done), $\phi$ is the phase difference across the junction and $J$ is the Josephson current (proportional to the current of Cooper-pairs across the junction). The ratio between the Josephson energy $E_J=h J/(2e)$ and the charging energy $E_C=(2e)^2/2C$ sets the third and last unitless parameter of the problem. The regime of $E_J/E_C\ll1$ is often termed ``weak coupling'' and is realized in ultra-small junctions (where the capacitance is strongly reduced, due to the small area), while $E_J/E_C\gg1$ is termed ``strong coupling''.

The circuit of Fig.\ref{fig:circuit} can also be represented in terms of the quantum Hamiltonian\cite{IngoldNazarov,Devoret}
\be H =  \frac{(2e)^2q^2}{2C} +(2e) q V(t) - E_J\cos(\phi) + H_R[\phi] \label{eq:H}\ee
Here the charge $q$ is cononically conjugated to $\phi$. The last term $H_R$ models the resistor as an infinite set of Harmonic oscillators. If the bath is  initially prepared in the ground state, the corresponding degrees of freedom can be exactly integrated-out, leading to a real-time action with non-local kernels. To avoid this complication, here we will work directly with the original circuit, rather than with its equivalent Hamiltonian representation.

\section{Renormalization of the Josephson coupling}\label{sec:first}

As a first step, let us consider the weak coupling regime $E_J\ll E_c$ and treat the Josephson coupling in a perturbative manner. If $E_J=0$ ($J=0$) the fluctuations of the system are given by the linear sum of two terms (1) Johnson-Nyquist noise, due to the equilibrium fluctuations on the resistor; (2) external noise. In what follows we will assume that the two sources are statistically independent. This is indeed a good approximation if the source of dissipation and the source of external noise are spatially separated (as in the case of Ref.[\onlinecite{Barends}]).

The spectral properties of the equilibrium fluctuations of a resistor are well known and equal to $\av{V_\w V^*_\w} = R \hbar \w {\rm cotgh}(\hbar\w/2T)$, where $T$ is the temperature of the resistor. For later reference we note that, if $T>0$, the low frequency tail of this spectrum is $\av{V_\w V^*_\w} = 2 R T$, giving rise to white noise (delta-function correlated in time) commonly associated with classical thermal noise. In the zero temperature limit the spectrum becomes
\be \av{V^*_\w V_\w} = R\hbar |\w| \label{eq:VVeq}\ee
This specific form of the spectrum, often called 	``quantum noise'', is a non-analytic function of $\w$, giving rise to long-tailed correlations in time\footnote{Recall that the derivatives of a function correspond to the momenta of its Fourier transform: if a function is non analytic, its Fourier transform must have long tails}.

\begin{figure*}
\centering
\includegraphics[scale=1]{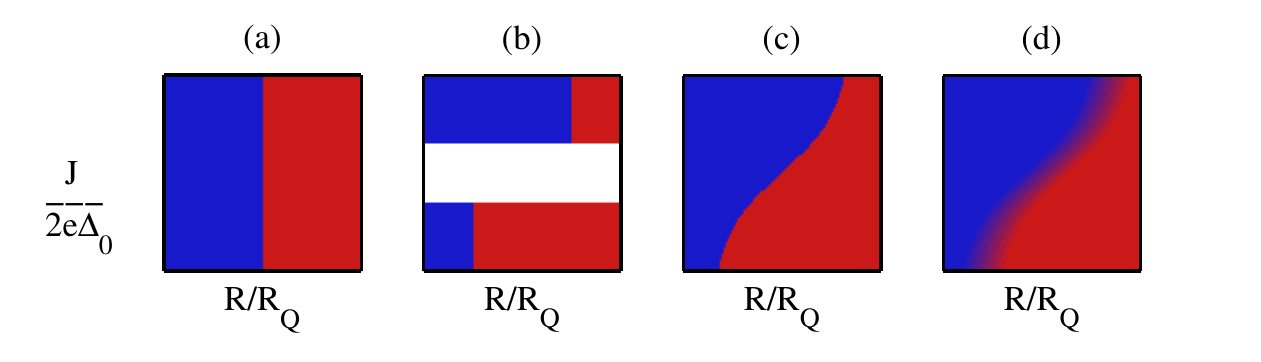}
\caption{Schematic phase diagram of the noisy shunted Josephson junction: blue areas are superconducting and red areas are insulating. (a) Equilibrium phase diagram displaying a quantum phase transition at $R=R_Q$. (b) First order corrections: the noise shifts the position of the transition in opposite directions, depending on the ratio between the Josephson coupling $J$ and the ultraviolet cutoff $\Delta_0=1/RC$. The white area indicates the region where higher order corrections are expect to play a dominant role. (c) Second order corrections to the voltage response: the noise renormalizes the resistance of the junction and changes the slope of the transition line. (d) Second order corrections to the voltage fluctuations: the noise generates an effective temperature which transforms the transition into a universal crossover controlled by the non-equilibrium quantum critical point.}\label{fig:phdia}
\end{figure*}

We now move to the voltage fluctuations induced by the external $1/f$ noise. Perhaps surprisingly, we will find that the their spectrum has precisely the same frequency dependence as the quantum noise, giving rise to an interesting collaboration between classical and quantum noise sources. The origin of this behavior can be traced back to the RC circuit acting as a derivative of the incoming signal. \footnote{For instance, if the incoming voltage is a step function, the current through the resistor, and hence its voltage, has a short exponentially decaying pulse, similar to a delta function.} More precisely, according to the linear circuit theory, the voltage over the resistor is $V_\w = R I_\w = R/(R + i/\w C) V_{N,\w}$. For frequencies significantly lower than $1/RC$, one than obtains $V_\w \approx i (RC)\w V_{N,\w}$. This approximation corresponds, in the renormalization group language, to the choice of an ultra-violet cutoff $\D_0=1/RC$. For a $1/f$ spectrum the voltage fluctuations:
\be \av{V^*_\w V_\w} = (RC)^2\w^2\av{V^*_{N,\w} V_{N,\w}} = R^2 (2e)^2 \frac{F}{2\pi} |\w| \label{eq:VV}\ee

In the absence of the Jopsephson junction, the circui is linear, and the two sources of noise (equilibrium and non-equilibrium) simply sum-up:
\be \av{V^*_\w V_\w} = R\hbar|\w| + R^2 (2e)^2 \frac{F}{2\pi} |\w| \label{eq:VV1}\ee
Note that only the first component is multiplied by $\hbar$, highlighting its quantum origin, while the second is of completely classical origin. %For a lucky coincidence, however, the two noise sorce scale in the same manner as function of the frequency.

A note of caution is now in place. Up to this point we have considered $V$ as a classical field, giving rise to real correlations $\av{V(t)V(t')}$. To enable a quantum mechanical approach to the problem, one has to keep in mind that the expectation value $S(t-t')=\av{V(t)V(t')}$ has both a real and imaginary part. At equilibrium, these two quantities are related by the fluctuation-dissipation theorem. In the presence of the noise, this relation is violated and the two quantities need to be determined independently. The real part $S_{\rm Re}(t-t')=\half\av{V(t)V(t')+V(t')(t)}$ is related to the fluctuations in the system and corresponds to (\ref{eq:VV1}). The imaginary part, $S_{\rm Im}(t-t')=\frac{1}{2i}\av{[V(t),V(t')]}$, describes the response of the system to an external probe. For $J=0$ the system is linear and the response function is independent on the noise:
$S_{Im}(\w) = \w(1/R + i \w C)^{-1}$.

%If the Josephson coupling is weak, we can assume that it is initially disconnected from the rest of the circuit. In this case, the fluctuations of the voltage over the resistor are determined by the sum of two terms: Josephson-Nyquist noise (becoming ``quantum noise'' in the limit of zero temperature) and the $1/f$ noise. The former is given be $\av{V^*_\w V_\w} = R~\hbar\w~ {\rm cotgh}(\hbar\w/2T) \to R \hbar |\w|$. The latter is $\av{V^*_\w V_\w} = R^2 \av{I^*_{{\rm ext},\w} I_{{\rm ext},\w}} = R^2 \w^2 \av{Q^*_{{\rm ext},\w} Q^*_{{\rm ext},\w}} = R^2 F (2e)^2 |\w|$. This result highlights the peculiarity of $1/f$ charge noise, that is to generate voltage fluctuations with the same frequency spectrum as equilibrium quantum noise.

We are now in the position of adding back the Josephson coupling $J\cos(\phi)$. For this task we need to compute the statistics of the phase fluctuations across the junction. Using the  Josephson relation $V(t) = (\hbar/2e)\partial_t \phi(t)$ and (\ref{eq:VV1}) we obtain
\bea &&\av{\left(\phi(t)-\phi(t')\right)^2} =\frac{R}{R_Q} \left(1 + \frac{R^2}{R_Q^2} F \right)\int d\w e^{i\w(t-t')} \frac{1}{|\w|}\nn\\
 &&= 2\pi \frac{R}{R_Q} \left(1 + \frac{R}{R_Q}F\right) {\rm log}(1+\D^2_0(t-t')^2)\label{eq:phiphi} \eea
Here we introduced back the capacitance, through the cutoff frequency $\D_0 = 1/RC$, in order to avoide the divergence of (\ref{eq:phiphi}) at short times.

Using eq.(\ref{eq:phiphi}) we can try to estimate whether the Josephson junction is capable of locking the phase across the junction $\phi$ or not. Due to the weak dependence of the logarithm on its argument, we first estimate $\log(\D_0|t-t'|) \approx 1$ and obtain $\delta\phi^2 \approx R/R_Q(1+R/R_Q F)$. If the phase fluctuations are small, the Josephson coupling is able to localize the phase, driving the system towards a superconducting state. If, on the other hand, the phase fluctuations are large, the Josephson coupling will have nearly no effect. As a consequence, we may naively expect a transition at $\delta\phi^2=1$ or
\be \frac{R^*}{R_Q}\left(1+\frac{R^*}{R_Q}F\right) = 1\label{eq:Rstar} \ee

This handwaving argument can be substantiated through the powerful ideas of the renormalization group (RG) approach. In general, the goal of any RG method is to study the macroscopic behavior of a model by gradually integrating over microscopic units. In our case there is no spatial dependance and the RG consists of averaging over fast processes. These fast processes renormalize the tunneling coupling by ``scrambling'' the phase across the junction. To quantify this process, we formally split the phase $\phi$ into slow (s) and fast (f) components and average over the latter:
\be J e^{i\phi} \equiv J e^{i(\phi_s + \phi_f)} \to J\av{e^{-\half\delta\phi^2_f}} e^{i\phi_s} \equiv J_{\rm eff} e^{i\phi_s}\label{eq:Jeff}\ee
Here we used the property of Gaussian distributions, for which $\av{e^{iA}}=e^{-1/2\av{A^2}}$. 

We now introduce an arbitrary frequency scale $\Delta$ separating the fast from the slow processes. If we choose $\Delta \gg J/(2e)$, it is reasonable to assume that these fast processes are independent on the Josephson coupling and their correlations are just given by (\ref{eq:VV1}). We than obtain that
\bea 
\delta\phi_f^2 &\equiv& (\phi_f(t)-\phi_f(0))^2 = \frac{R}{R_Q}\left(1 + \frac{R}{R_Q} F \right)\int_{|\w|>\D}d\w\frac1{|\w|}\nn \\
&\approx&  \frac{R}{R_Q}\left(1 + \frac{R}{R_Q} F \right)\log(\D/\D_0)\label{eq:deltaphi2}
\eea
Combining (\ref{eq:Jeff}) and (\ref{eq:deltaphi2}) one obtains:
\be \frac{J_{\rm eff}}{\D} = \frac{J}{\D_0} \left(\frac{\D}{\D_0}\right)^{\frac{R}{R_Q}\left(1 + \frac{R}{R_Q} F \right)-1} \label{eq:scaleJ}\ee

Eq.(\ref{eq:scaleJ}) indicates the inset of a transition precisely at (\ref{eq:Rstar}). For $R>R^*$ the exponent in eq.(\ref{eq:scaleJ}) is positive. Hence, as we reduce $\D$ towards zero the ratio $J_{\rm eff}/\D$ tends to zero. In this case, the Josephson coupling is said to be ``irrelevant'' and the junction behaves as an insulator. In the opposite case ($R<R^*$) the exponent in (\ref{eq:scaleJ}) is negative and the renormalized Josephson coupling grows, inducing a superconducting behavior.

\begin{figure*}
\includegraphics[scale=0.7]{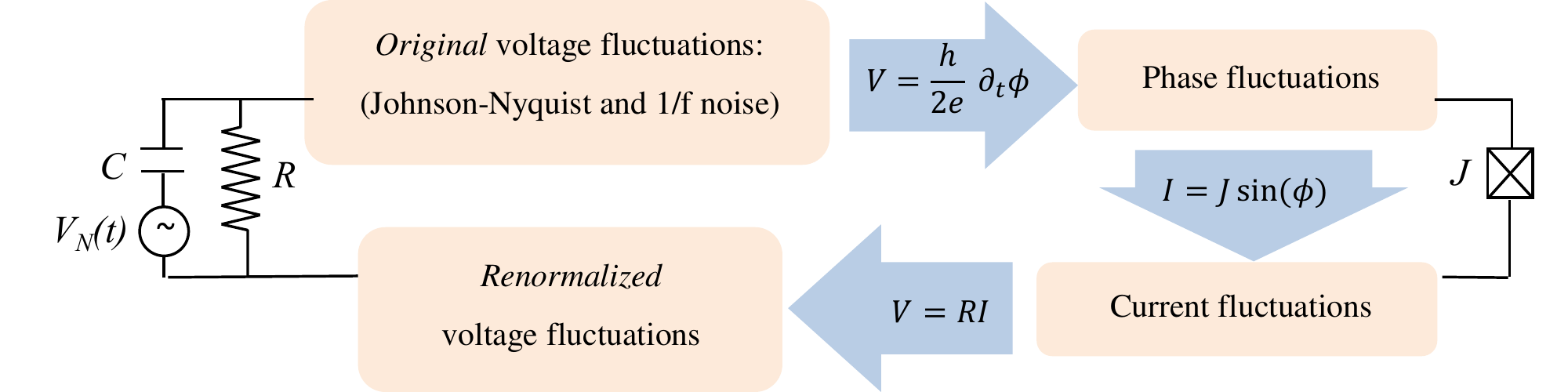}
\caption{Flow diagram of the feed-back loop generating a renormalization of the voltage fluctuations and the generation of an effective temperature (see text)}\label{fig:feedback}
\end{figure*}

Until this point we considered only the weak coupling limit and performed a perturbative expansion in $E_J/E_c\ll1$. We will now see that similar arguments apply to the strong coupling limit $E_J/E_c\gg1$ as well. In this case, we can assume that the voltage across the junction is zero. The current through the junction is given by two terms: quantum noise and external noise. In analogy to the previous analysis we find
\be \av{I_\w I^*_\w} = \left(\frac{\hbar}{R} + F (2e)^2\right)|\w| \ee
Or, in terms of charge fluctuations,
\be \av{Q(t)Q(t')} = (2e)^2\frac{R_Q}{R}\left(1 + \frac{R}{R_Q}F \right){\rm log}\left(\frac{t-t'}{\tau}\right) \ee
From the above result we can guess the existance of a charge localization phase transition. If the charge fluctuations are larger than the charge of a single Cooper pair $(2e)^2$, the junction behaves a superconductor. If the charge fluctuations are smaller, the capacitance becomes dominant and the junction behaves as an insulator. %Formally, one can show that the charging energy of the junction corresponds to a periodic function of $q/2e$, often referred to as ``dual coupling''. 
Indeed, an exact duality-transformation\cite{FisherZwerger,us-prb} shows that the system undergoes a quantum phase transition at:
\be \frac{R_Q}{R^*} \left(1 + \frac{R^*}{R_Q}F\right) = 1\label{eq:Rstar2}\ee

Equation (\ref{eq:Rstar}) and (\ref{eq:Rstar2}) show the first non-trivial effect of the noise, namely the shift of the transition away from its universal equilibrium value $R=R_Q$. This effect is schematically shown in Fig.\ref{fig:phdia}(b). Note that in the weak coupling limit the critical resistance is smaller than $R_Q$, indicating that the insulating regime is stabilized by the noise. On the other hand, in the strong coupling limit, the critical resistance $R^*$ is larger than $R_Q$ and the superconductor is stabilized. Our intuitive approach makes evident the origin of this effect: the noise increases the fluctuations of both the phase and the charge, always stabilizing the delocalized phase (i.e. the insulator at weak coupling and the superconductor at strong coupling). The same would be true for pure thermal fluctuations, which however would not change the critical correlations accordingly.

\section{Response function and renormalization of the resistance}\label{sec:second}
The second major effect discovered in Ref.[\onlinecite{us-prb}] is the generation of an effective temperature. To understand the origin of this effect and its intuitive meaning, we need to consider higher order processes in the Josephson coupling, related to feed-back effects. As discussed above, the noisy $RC$ circuit generates (non-equilibrium) voltage fluctuations with $|\omega|$ spectrum, which translate into phase fluctuations, according to the Josephson {\it voltage} law. Then, following the Josephson {\it current} law, the phase fluctuations translate into current fluctuations which are fed back into the RC circuit. When these current fluctuations pass through the resistor, they generates additional voltage fluctuations that correct the original voltage spectrum, and so on so forth (See Fig.{\ref{fig:feedback}).

To second order in the Josephson coupling we then have
\bea \delta S(t-t') &=& R^2 J^2 \av{\sin(\phi(t))\sin(\phi(t'))} \nn\\
&=& R^2 J^2 e^{-\half\av{(\phi(t)-\phi(t'))^2}} \label{eq:VV2}\eea
As before, $\delta S(t-t')$ has both a real and an imaginary part, which renormalize respectively the fluctuations and the response of the system. To identify these two components it is useful to define two real functions, $C(t)={\rm Re}\av{(\phi(t)-\phi(0))^2}$ and $R(t)={\rm Im}\av{(\phi(t)-\phi(0))^2}=\frac{1}{2i}\av{[\phi(t),\phi(t')]}$, corresponding respectively to the correlations and response of $\phi(t)$.  For the specific choice of the cutoff introduced in Ref. \cite{us-prb}, $C(t)$ is given by (\ref{eq:phiphi}) and $R(t)$ by $R/R_Q {\rm atan}(t/\D_0)$. Using these definitions (\ref{eq:VV2}) we immediately obtain $\delta S(t)=\delta S_{\rm Re}(t) + i \delta S_{\rm Im}(t)$, where
\bea 
\delta S_{\rm Re}(t) &=& R^2 J^2\cos(R(t)) e^{- C(t)} \label{eq:SRe}\\
\delta S_{\rm Im}(t) &=& R^2 J^2\sin(R(t)) e^{- C(t)}\label{eq:SIm}
\eea

Let us first consider the contribution to the response of the junction $\delta S_{\rm Im}(t)$. The Fourier transform of this function is plotted in Fig. \ref{fig:allinone}(a) at equilibrium (solid curve) and in the presence of a strong $1/f$ noise (dashed curve).  The comparison between the two curves shows a remarkable difference. At equilibrium the slope of the curve tends to zero as $\w\to0$, while in the latter it tends to a constant. This trend is highlighted in Fig.\ref{fig:allinone}(d), where the zero-frequency limit of the derivative of $\delta S_{\rm Im}(\w)$ is shown as function of the noise strength $F$. At low frequencies we can  approximate the contribution to the response of the system as $\delta S_{\rm Im} \approx \w \partial_\w S_{\rm Im}(\w=0) $. This term sums-up to the pre-existing (bare) response $S_{\rm Im} \approx R \w$. We conclude that the resistance is renormalized by
\be dR(J) = \partial_\w S_{\rm Im}(\w=0)\label{eq:dR} \ee

The non-equilibrium renormalization of the resistance has important effects on the resulting phase diagram. Due to this term, the transition is now located at $R + dR(J) = R^*$. Hence, with increasing $J$, the transition moves to higher values of $R$ (as shown in Fig.\ref{fig:allinone}(b), the correction is negative $dR(J)<0$). When the Josephson coupling becomes of the order of the cutoff frequency ($J/2e \approx \D_0$) the transition is moved back to $R\approx R_Q$. In the strong coupling regime, one should apply the duality transformation described above. This analysis  shows that, starting from $J\to\infty$, when $J$ is decreased the transition moves to lower values of $R$. The weak and strong coupling limit therefore lead to a consistent picture, shown in Fig.\ref{fig:phdia}(c).

\begin{figure}
\includegraphics[scale=0.9]{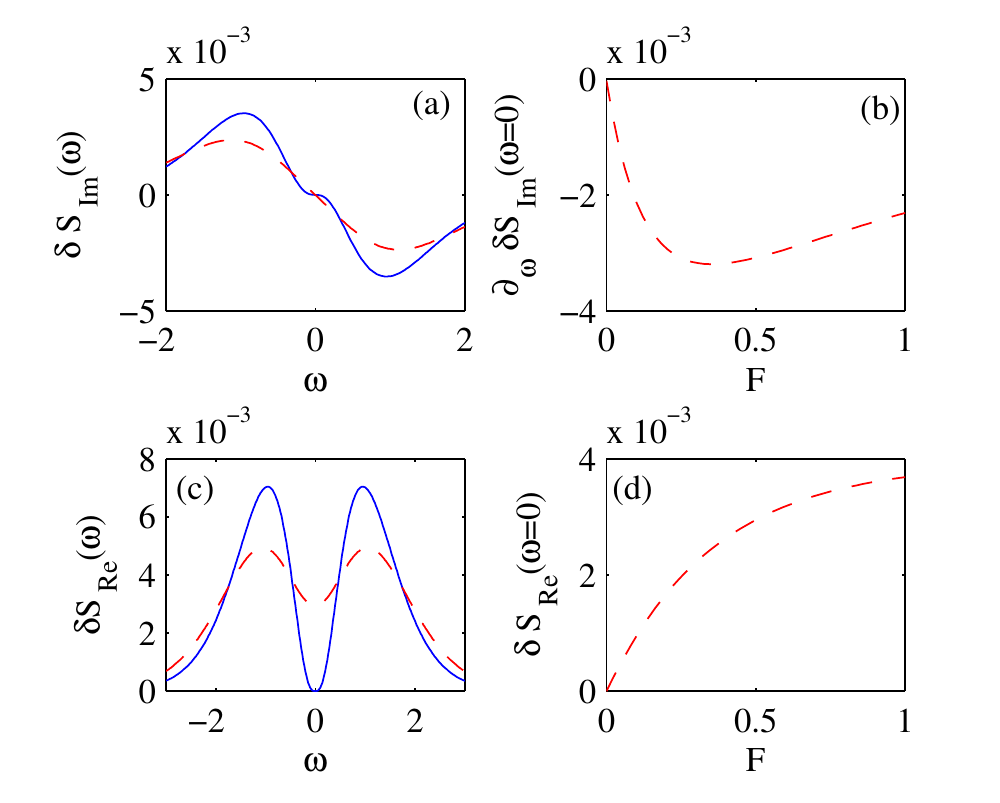}
\caption{Second order contributions: (a) Renormalization of the voltage fluctuations $\delta S_{\rm Re}(\w)$ at equilibrium (solid curve, $R/R_Q=2$, $g/\D_0=0.1$) and in the presence of $1/f$ noise (red dashed curve, $R/R_Q=2$, $F=0.5$, $g/D_0=0.1$). (b) Zero frequency component as function of the noise strength $F$. (c) Renormalization of the response function $\delta R_{\rm Im}(\w)$ at equilibrium (solid) and in the presence of $1/f$ noise with $F=0.5$ (dashed), for $R/R_Q=2$. (d) Zero frequency derivative as function of the noise strength $F$.  All axes are given in units of the cutoff frequency $\Delta_0=1$.}\label{fig:allinone}
\end{figure}

\section{Voltage fluctuations and effective temperature} \label{sec:third}

We now move to the third main effect of the external noise, namely the generation of an effective temperature. For this task, we need to consider the contributions of the Josephson coupling to the voltage fluctuations of the junction $S_{\rm Re}(\w)$. The relevant expression is given in (\ref{eq:SRe}) and depicted in Fig.\ref{fig:allinone}(c). We note a significant difference between the low-frequency behavior at equilibrium and in the presence of $1/f$ noise. The equilibrium curve tends to zero at zero frequency, while the non-equilibrium curve has a finite zero frequency component. This feature is highlighted in Fig. \ref{fig:allinone}(b), where the value of the zero frequency component of $\delta S_{\rm Re}$ is shown as function of the noise strength $F$. 

The zero frequency component of $\delta S_{\rm Re}$ is the origin of the finite effective temperature. As we discussed in the introduction, a finite temperature  corresponds to white noise with spectrum $S_{\rm Re} = R T$. Here the spectrum is highly non linear, but at low enough frequencies, we can neglect the bare term $\sim |\w|$ and approximate $S_{\rm Re}(\w) \approx \delta S_{\rm Re}(\w=0)$ to obtain:
\be T_{\rm eff} = \frac1R \delta S_{\rm Re}(\w=0)\label{eq:Teff} \ee
It is worth noting that this effective temperature has to be understood only in an {\it RG sense}. It only determines the behavior of low-frequency behavior of the junction, while the high-frequency behavior is still strongly out of equilibrium.

This effective temperature has drastic effects on the phase diagram, transforming the sharp phase transition into a smooth crossover. As we explained above, the first order predictions of a sharp transition were based on the power-law dependence of the renormalized Josephson coupling, eq. (\ref{eq:scaleJ}). However, as we now understand, this equation is valid only at frequency scales larger than $T_{\rm eff}/\hbar$. At this scale the thermal noise sets-in and leads to an exponential decay of the effective Josephson coupling. For the superconducting behavior to be observable one has to require  the renormalized Josephson coupling at the scale $\D=T_{\rm eff}$ to be larger than the effective temperature itself, or:
\be \frac{J}{\D_0} > \left(\frac{T_{\rm eff}}{\D_0}\right)^{1-\frac{R}{R_Q}(1+\frac{R}{R_Q}F)} \label{eq:cross}\ee
Recalling that $T_{\rm eff}\sim J^2$ we find that, if $R/R_Q(1+R/R_QF)<1/2$, the condition (\ref{eq:cross}) is always satisfied for $J\to 0$. In the intermediate regime $1/2<R/R_Q(1+R/R_Q F)<1$, on the other hand, a finite $J$ is needed to obtain a superconductor. This effect is depicted schematically in Fig.\ref{fig:phdia}(d).

\section{Renormalization of the current-voltage characteristic}\label{sec:IVcurve}
In the previous sections we described the effects of the noise on the renormalization of the model and on the resulting phase diagram. Here, we will describe how these effects can be probed by measuring the non-linear current-voltage characteristic of the model. Time dependent perturbation theory\cite{us-prb} shows that the difference between the total current in the junction and the current passing through the resistor $I_s=I-V/R$ is given by:
\bea I - \frac{V}R &=& J^2 \int dt~\av{[\cos\left(\phi(t)+\frac{2eV t}h\right),\cos(\phi(0))]}\nn \\
&=&  J^2 \int dt~e^{i(2e)V t/h} {\rm Im}\av{\cos(\phi(t))\cos(\phi(0))}\nn\\
&=& \frac{1}{R^2} \delta S_{\rm Im}(\w=2eV/h)\label{eq:Is}\eea
Here we used the definition of $\delta S_{\rm Im}$ given in eq.(\ref{eq:SIm}).

Let us now study the behavior of this function in different regimes. For small voltages, we can expand $\delta S_{\rm Im}$ in Taylor series of the frequency to obtain
\be  I = \frac{V}{R}\left[1 - \frac 1{R}\partial_\w\delta S_{\rm Im}(\w=0)\right] \ee
Inverting this expression we obtain:
\be \partial_\w\delta S_{\rm Im}(\w=0) =  \frac{R I - V}{V/R} \approx \frac{\delta V}{I} = \delta R. \ee
This identity, already given in Sec.\ref{sec:second}, acquires now a clearer significance: the low-frequency slope of $\delta S_{\rm Im}$ corresponds to a renormalization of the low-voltage resistance.

\begin{figure}
\includegraphics[scale=0.9]{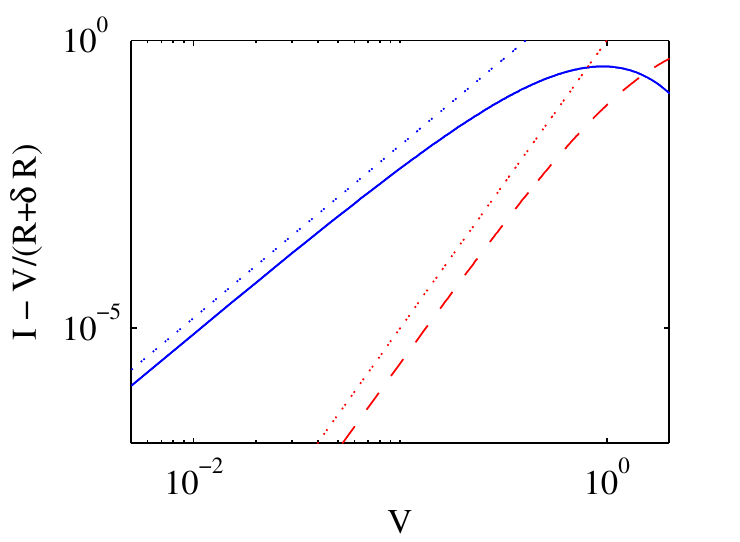}
\caption{Non linear current-voltage characteristic, given by the difference between the supercurrent and the (renormalized) normal current $I-V/(R+\delta R)$, according to eq.(\ref{eq:nonlin}): equilibrium result (solid curve, $R/R_Q=2$) and non-equilibrium result (dashed curve, $R/R_Q=2$ and $F=0.5$). The dotted lines represent the scaling result $ \sim V^{2R/R_Q-1}$.}
\label{fig:IVcurve}\end{figure}

At larger voltages the junction deviates from this linear slope according to
\be I - \frac{V}{R+\delta R} = \frac1{R^2}\left[S_{\rm Im}(\w) - \w\partial_\w S_{\rm Im}(\w)\right]\label{eq:nonlin}\ee 
This quantity is plotted in Fig.\ref{fig:IVcurve} on a log-log scale and clearly displaying an algebraic dependance. To understand this behavior, we consider the long-time limit of $S_{\rm Im}(t) \sim t^{-2R/R_Q(1+F R/R_Q)}$, leading to $S_{\rm Im}(t) \sim \w^{2R/R_Q(1+F R/R_Q)-1}$, or $I - V/(R+\delta R) \sim V^{2R/R_Q(1+F R/R_Q)-1}$. This power-law dependance is a direct consequence of the renormalization of the Josephson coupling discussed in Sec.\ref{sec:first}.

Note that the non-linear behavior precisely disappears at the phase transition, where eq. (\ref{eq:nonlin}) gives $I \sim V$. Thus, measuring the non-linear IV curve allows to determine the precise position of the phase transition. This conclusion is however modified when one takes into account the presence of a finite effective temperature. As discussed in Sec.\ref{sec:third}, the predicted power-law behavior will terminate at the frequency scale $\w = T_{\rm eff}$. This poses a limitation on our capability of distinguishing between the different phases, thus transforming the sharp phase transition into a smooth crossover.

\section{Summary and discussion}\label{sec:summary}

In this paper we consider the noise-driven resitively-shunted Josephson junction, originally proposed in Ref.[\onlinecite{us-nature}] and [\onlinecite{us-prb}]. With respect to these two works, here we focus on the intuitive derivation and understanding of the results. Following the ideas proposed in Ref.[\onlinecite{LeggettReview}], we present the first order results of the renormalization group as a time-dependent average over fast processes. For the second order processes, involving the renormalization of the temperature and the resistance, we present even simpler calculations, based on time-dependent perturbation theory. 

The resulting non-equilibrium effects significantly modify the resulting phase diagram, as probed by the current-voltage characteristic of the junction. The renormalization of the resistance affects the slope at the low current limit of the curve. The renormalization of the Josephson junction determines the non-linear behavior at larger voltages, and the effective temperature sets the transition frequency between these two regimes.

One important effect which we did not consider here (nor in any of our previous papers) are the deviations from $1/f$ noise. It is known that experimental spectra always deviate from this theoretical curve. Based on dimensional analysis it is natural to distinguish between noise sources affecting the long-time behavior of the correlations (relevant noise) and those which leave it unchanged (irrelevant). Without pretending to study this problem in depth, we have computed numerically the effective temperature and the renormalization of the resistance for specific cases of relevant and irrelevant noise sources.

\begin{figure}[t]
\centering
\includegraphics[scale=0.8]{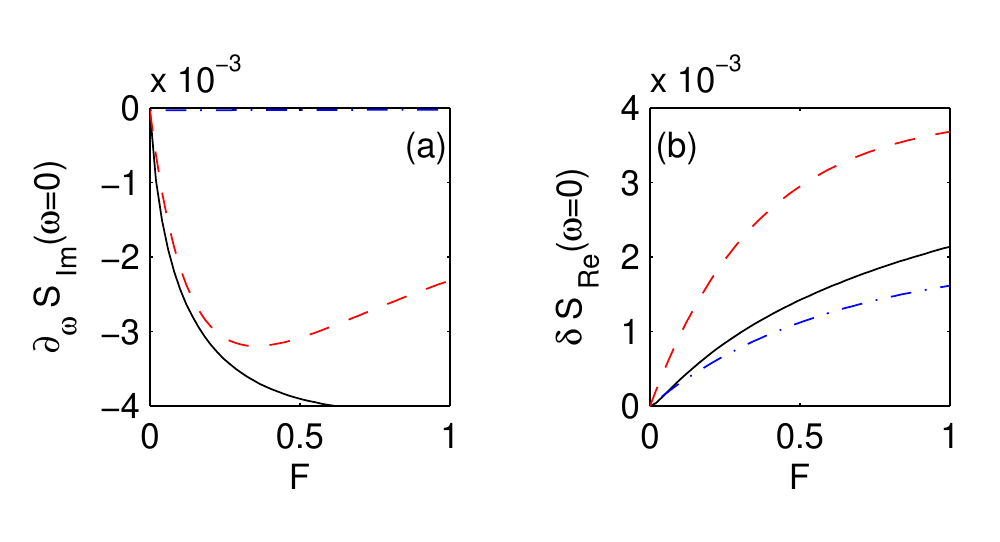}
\caption{Effects of relevant and irrelevant noise sources, for $R/R_Q=2$. (i) Dashed curve: $1/f$ noise. (ii) Solid curve: a relevant noise source, entering the phase correlations through $C(t) = \frac{R}{R_Q}\log\left(1+\D^2_0 t^2 + F \D_0^4 t^4\right)$. (iii) Dashed-dotted curve: an irrelevant noise source entering the phase correlations through $C(t) =\frac{R}{R_Q} \log\left(1+\D_0^2 t^2 +F\frac1{\D_0} |t|\right)$.}\label{fig:allinone2}
\end{figure}

In general, we found that the renormalization of the temperature is always present, independent of the relevance of the noise, in agreement with the general arguments given in Sec.\ref{sec:third}. The renormalization of the resistance, on the other hand, strongly depends on the relevance of the noise. Two specific examples are given in Fig.\ref{fig:allinone2}. In the case of a relevant noise source (solid curve) the renormalization of the dissipation is extremely large even for pretty low noise strengths. In the case of an irrelevant source (dotted-dashed curve), on the other hand, the renormalization of the dissipation is negligible. A complete understanding of these effects is still lacking.

We thank S. Huber, D. Huse, A. Mitra and A. Rosch for stimulating discussions. This research was supported in part by the US israel binational science foundation (EA and ED), the Israel Science foundation (EA), and the Swiss NSF under MaNEP and Division II (TG).

\appendix
\section{Experimental realization with ultracold atoms}
In this appendix we discuss two possible ways to realize and probe the noisy shunted Josephson junction (\ref{eq:H}), using ultracold atoms confined to one dimension (See Fig.\ref{fig:ultracold}). In both realizations the dissipative bath corresponds to the low-energy excitations (phonons) of the liquid. In one dimension the phonons have universal properties\cite{Haldane}: their spectrum is linear and their spectral density constant. When integrated out, these modes precisely correspond to a linear resistor\cite{KaneFisher}. 

The easiest way to understand the mapping between the 1d system and the resistor is to compute the correlation function of the displacement field $\theta(x,t)$ in the 1d model and to compare them with the (equilibrium) correlations of the phase across the junction $\phi(t)$. If the phonons are prepared in their ground state, the correlations are:
\bea \av{\theta(x,t)\theta(x,t')} &=& \sum_q \av{\theta^2_q}\cos(\w_k t)\label{eq:LL}\\
 &=& K \int_{-\pi/a}^{\pi/a} dq~\frac{1}{|q|}\cos(c q t) \approx K \log\left(\frac{c t}{a}\right)\nn\eea
Here $\w_q = c q$ is the spectrum of the phonons and $a$ the average atomic distance (which acts as a UV cutoff). In the second row we used the fact that the zero-point motion of a harmonic oscillator is proportional to the inverse of its eigenfrequency and we introduced the proportionality constant $K$. For the proper choice of units, both the field $\theta$ and the parameter $K$ are unitless. We refer the reader to Ref.\onlinecite{Giamarchi} for a detailed derivation of the relation between $K$ (the ``Luttinger parameter'') and the microscopic parameters of the models. At equilibrium (\ref{eq:LL}) and (\ref{eq:phiphi}) coincide, provided that we identify $K\to R/R_Q$ and $\theta(x,t)\to \phi(t)$.

\begin{figure}
\centering
\includegraphics[scale=0.8]{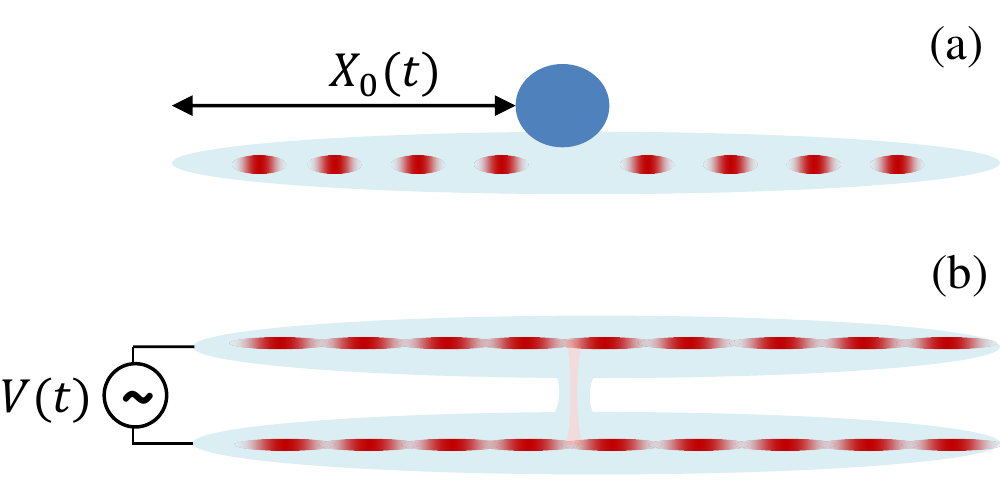}
\caption{Experimental realization of the noisy shunted Josephson junction with ultracold atoms confined to one dimension. (See text.)}\label{fig:ultracold}
\vspace{-0.5cm}
\end{figure}

We now move to the physical realization of the Josephson coupling $J$. Following the pioneering work by Kane and Fisher\cite{KaneFisher}, our first proposed realization consists of a local impurity weakly coupled to the one dimensional system. As known from the literature\cite{Giamarchi}, the energy associated with a (back-scattering) impurity at position $X_0$ is $H_{\rm impurity} = V a \rho(x=X_0) = V \cos(2\theta + 2\pi X_0/a)$. To drive the system out of equilibrium we propose to stochastically shift the position of the impurity as function of time, with $1/f$ spectrum, such that $\av{X_0(\w)X^*_0{(\w)}}= (a/\pi)^2 F/|\w|$. To make a direct connection with the noisy Josephson junction (\ref{eq:H}) it is enough to define a new variable $\phi(t) =\half\left(\theta(t) -\pi X_0(t)/a\right)$. For this coordinate the bare fluctuations are given by the sum of an equilibrium ($\sim K/|\w|$) and a non-equilibrium ($\sim F/|\w|$) component and the non linear coupling ($\cos(\phi)$) is time independent.

Having established a formal equivalence between the noisy shunted Josephson junction and a vibrating impurity in a one dimensional liquid, we now describe the physical consequences of this equivalence. For a shunted Josephson junction the natural physical quantity to look at is the non-linear I-V curve. In the proposed realization, a finite current bias $I = \dot Q$ can be induced by dragging the impurity at a constant velocity (in addition to the random $1/f$ fluctuations), $\bar X_0(t) = a I/(2e)$. The supercurrent $I_s \sim \sin (\phi)$ can be probed by measuring the atomic density at the distance $a/2$ from the impurity $\rho(x=X_0+a/2) = \cos(2\phi + \pi X_0(t)/a + \pi/2)$, thus giving access to the non-linear I-V curve of the junction.

Our second proposed realization is depicted in Fig.\ref{fig:ultracold}(b). Here the model is realized in the anti-symmetric modes of two parallel 1d liquids. If the tunneling between the tubes is allowed only at a given position, it immediately maps into a Josephson coupling $V_{\rm tunneling}= t\cos(\delta\phi)$, where $\delta\phi=\phi_{1}-\phi_2$ is the phase difference between the two tubes and $t/\hbar$ the tunneling rate. To introduce time-dependent noise in the system, we propose to apply a time-dependent potential-difference between the tubes, and to define accordingly $\phi =\delta\phi+\int dt~V(t)$. With respect to the previous realization, the voltage difference appears in the effective field with an additional integral over time. Thus, to mimic the $1/f$ charge noise, we need to consider a voltage spectrum $\av{V^*(\w)V(\w)}\sim|\w|$. In fact, this type of noise is easier to generate than $1/f$ noise because its correlations decay algebraically rather than logarithmically.

In this second realization, we can mimic a constant current bias by applying an additional DC potential difference between the two tubes. The supercurrent can be probed by measuring the interference fringes of the two condensates (at the position of the tunneling junction), on the lines of Ref.\onlinecite{Schmiedmayer2}. For high voltage differences the interference fringes completely disappear, corresponding to a linear I-V curve. As we lower the voltage, the fringes are expected to slowly reappear, indicating a non-linear I-V curve of the original model.

%\bibliographystyle{aipnum4-1}

%\bibliography{noneq}

\begin{thebibliography}{25}%
\makeatletter
\providecommand \@ifxundefined [1]{%
 \@ifx{#1\undefined}
}%
\providecommand \@ifnum [1]{%
 \ifnum #1\expandafter \@firstoftwo
 \else \expandafter \@secondoftwo
 \fi
}%
\providecommand \@ifx [1]{%
 \ifx #1\expandafter \@firstoftwo
 \else \expandafter \@secondoftwo
 \fi
}%
\providecommand \natexlab [1]{#1}%
\providecommand \enquote  [1]{``#1''}%
\providecommand \bibnamefont  [1]{#1}%
\providecommand \bibfnamefont [1]{#1}%
\providecommand \citenamefont [1]{#1}%
\providecommand \href@noop [0]{\@secondoftwo}%
\providecommand \href [0]{\begingroup \@sanitize@url \@href}%
\providecommand \@href[1]{\@@startlink{#1}\@@href}%
\providecommand \@@href[1]{\endgroup#1\@@endlink}%
\providecommand \@sanitize@url [0]{\catcode `\\12\catcode `\$12\catcode
  `\&12\catcode `\#12\catcode `\^12\catcode `\_12\catcode `\%12\relax}%
\providecommand \@@startlink[1]{}%
\providecommand \@@endlink[0]{}%
\providecommand \url  [0]{\begingroup\@sanitize@url \@url }%
\providecommand \@url [1]{\endgroup\@href {#1}{\urlprefix }}%
\providecommand \urlprefix  [0]{URL }%
\providecommand \Eprint [0]{\href }%
\providecommand \doibase [0]{http://dx.doi.org/}%
\providecommand \selectlanguage [0]{\@gobble}%
\providecommand \bibinfo  [0]{\@secondoftwo}%
\providecommand \bibfield  [0]{\@secondoftwo}%
\providecommand \translation [1]{[#1]}%
\providecommand \BibitemOpen [0]{}%
\providecommand \bibitemStop [0]{}%
\providecommand \bibitemNoStop [0]{.\EOS\space}%
\providecommand \EOS [0]{\spacefactor3000\relax}%
\providecommand \BibitemShut  [1]{\csname bibitem#1\endcsname}%
\let\auto@bib@innerbib\@empty
%</preamble>
\bibitem [{\citenamefont {Calabrese}\ and\ \citenamefont
  {Cardy}(2007)}]{Cardy}%
  \BibitemOpen
  \bibfield  {author} {\bibinfo {author} {\bibfnamefont {P.}~\bibnamefont
  {Calabrese}}\ and\ \bibinfo {author} {\bibfnamefont {J.}~\bibnamefont
  {Cardy}},\ }\href {http://stacks.iop.org/1742-5468/2007/i=06/a=P06008}
  {\bibfield  {journal} {\bibinfo  {journal} {Journal of Statistical Mechanics:
  Theory and Experiment}\ }\textbf {\bibinfo {volume} {2007}},\ \bibinfo
  {pages} {P06008} (\bibinfo {year} {2007})}\BibitemShut {NoStop}%
\bibitem [{\citenamefont {De~Grandi}, \citenamefont {Gritsev},\ and\
  \citenamefont {Polkovnikov}(2010)}]{DeGrandi}%
  \BibitemOpen
  \bibfield  {author} {\bibinfo {author} {\bibfnamefont {C.}~\bibnamefont
  {De~Grandi}}, \bibinfo {author} {\bibfnamefont {V.}~\bibnamefont {Gritsev}},
  \ and\ \bibinfo {author} {\bibfnamefont {A.}~\bibnamefont {Polkovnikov}},\
  }\href {\doibase 10.1103/PhysRevB.81.012303} {\bibfield  {journal} {\bibinfo
  {journal} {Phys. Rev. B}\ }\textbf {\bibinfo {volume} {81}},\ \bibinfo
  {pages} {012303} (\bibinfo {year} {2010})}\BibitemShut {NoStop}%
\bibitem [{\citenamefont {Mondal}, \citenamefont {Sengupta},\ and\
  \citenamefont {Sen}(2009)}]{Sengupta}%
  \BibitemOpen
  \bibfield  {author} {\bibinfo {author} {\bibfnamefont {S.}~\bibnamefont
  {Mondal}}, \bibinfo {author} {\bibfnamefont {K.}~\bibnamefont {Sengupta}}, \
  and\ \bibinfo {author} {\bibfnamefont {D.}~\bibnamefont {Sen}},\ }\href
  {\doibase 10.1103/PhysRevB.79.045128} {\bibfield  {journal} {\bibinfo
  {journal} {Phys. Rev. B}\ }\textbf {\bibinfo {volume} {79}},\ \bibinfo
  {pages} {045128} (\bibinfo {year} {2009})}\BibitemShut {NoStop}%
\bibitem [{\citenamefont {Diehl}\ \emph {et~al.}(2008)\citenamefont {Diehl},
  \citenamefont {Micheli}, \citenamefont {Kantian}, \citenamefont {Kraus},
  \citenamefont {Buchler},\ and\ \citenamefont {Zoller}}]{DiehlZoller}%
  \BibitemOpen
  \bibfield  {author} {\bibinfo {author} {\bibfnamefont {S.}~\bibnamefont
  {Diehl}}, \bibinfo {author} {\bibfnamefont {A.}~\bibnamefont {Micheli}},
  \bibinfo {author} {\bibfnamefont {A.}~\bibnamefont {Kantian}}, \bibinfo
  {author} {\bibfnamefont {B.}~\bibnamefont {Kraus}}, \bibinfo {author}
  {\bibfnamefont {H.~P.}\ \bibnamefont {Buchler}}, \ and\ \bibinfo {author}
  {\bibfnamefont {P.}~\bibnamefont {Zoller}},\ }\href@noop {} {\bibfield
  {journal} {\bibinfo  {journal} {Nature Physics}\ }\textbf {\bibinfo {volume}
  {4}},\ \bibinfo {pages} {878} (\bibinfo {year} {2008})}\BibitemShut {NoStop}%
\bibitem [{\citenamefont {Prosen}\ and\ \citenamefont
  {Pi\ifmmode~\check{z}\else \v{z}\fi{}orn}(2008)}]{Prosen}%
  \BibitemOpen
  \bibfield  {author} {\bibinfo {author} {\bibfnamefont {T.~c.~v.}\
  \bibnamefont {Prosen}}\ and\ \bibinfo {author} {\bibfnamefont
  {I.}~\bibnamefont {Pi\ifmmode~\check{z}\else \v{z}\fi{}orn}},\ }\href
  {\doibase 10.1103/PhysRevLett.101.105701} {\bibfield  {journal} {\bibinfo
  {journal} {Phys. Rev. Lett.}\ }\textbf {\bibinfo {volume} {101}},\ \bibinfo
  {pages} {105701} (\bibinfo {year} {2008})}\BibitemShut {NoStop}%
\bibitem [{\citenamefont {Ringel}\ and\ \citenamefont
  {Gritsev}(2012)}]{Gritsev}%
  \BibitemOpen
  \bibfield  {author} {\bibinfo {author} {\bibfnamefont {M.}~\bibnamefont
  {Ringel}}\ and\ \bibinfo {author} {\bibfnamefont {V.}~\bibnamefont
  {Gritsev}},\ }\href {http://stacks.iop.org/0295-5075/99/i=2/a=20012}
  {\bibfield  {journal} {\bibinfo  {journal} {EPL (Europhysics Letters)}\
  }\textbf {\bibinfo {volume} {99}},\ \bibinfo {pages} {20012} (\bibinfo {year}
  {2012})}\BibitemShut {NoStop}%
\bibitem [{\citenamefont {Mitra}\ \emph {et~al.}(2006)\citenamefont {Mitra},
  \citenamefont {Takei}, \citenamefont {Kim},\ and\ \citenamefont
  {Millis}}]{MitraMillis}%
  \BibitemOpen
  \bibfield  {author} {\bibinfo {author} {\bibfnamefont {A.}~\bibnamefont
  {Mitra}}, \bibinfo {author} {\bibfnamefont {S.}~\bibnamefont {Takei}},
  \bibinfo {author} {\bibfnamefont {Y.~B.}\ \bibnamefont {Kim}}, \ and\
  \bibinfo {author} {\bibfnamefont {A.~J.}\ \bibnamefont {Millis}},\ }\href
  {\doibase 10.1103/PhysRevLett.97.236808} {\bibfield  {journal} {\bibinfo
  {journal} {Phys. Rev. Lett.}\ }\textbf {\bibinfo {volume} {97}},\ \bibinfo
  {pages} {236808} (\bibinfo {year} {2006})}\BibitemShut {NoStop}%
\bibitem [{\citenamefont {Mitra}\ and\ \citenamefont
  {Millis}(2008)}]{MitraMillis2}%
  \BibitemOpen
  \bibfield  {author} {\bibinfo {author} {\bibfnamefont {A.}~\bibnamefont
  {Mitra}}\ and\ \bibinfo {author} {\bibfnamefont {A.~J.}\ \bibnamefont
  {Millis}},\ }\href {\doibase 10.1103/PhysRevB.77.220404} {\bibfield
  {journal} {\bibinfo  {journal} {Phys. Rev. B}\ }\textbf {\bibinfo {volume}
  {77}},\ \bibinfo {pages} {220404} (\bibinfo {year} {2008})}\BibitemShut
  {NoStop}%
\bibitem [{\citenamefont {Feynman}\ and\ \citenamefont
  {Vernon}(1963)}]{FeynmanVernon}%
  \BibitemOpen
  \bibfield  {author} {\bibinfo {author} {\bibfnamefont {R.~P.}\ \bibnamefont
  {Feynman}}\ and\ \bibinfo {author} {\bibfnamefont {F.~L.}\ \bibnamefont
  {Vernon}},\ }\href {\doibase DOI: 10.1016/0003-4916(63)90068-X} {\bibfield
  {journal} {\bibinfo  {journal} {Annals of Physics}\ }\textbf {\bibinfo
  {volume} {24}},\ \bibinfo {pages} {118 } (\bibinfo {year}
  {1963})}\BibitemShut {NoStop}%
\bibitem [{\citenamefont {Cladeira}\ and\ \citenamefont
  {Leggett}(1983)}]{CaldeiraLeggett}%
  \BibitemOpen
  \bibfield  {author} {\bibinfo {author} {\bibfnamefont {A.~O.}\ \bibnamefont
  {Cladeira}}\ and\ \bibinfo {author} {\bibfnamefont {A.~J.}\ \bibnamefont
  {Leggett}},\ }\href@noop {} {\bibfield  {journal} {\bibinfo  {journal}
  {Physica A}\ }\textbf {\bibinfo {volume} {121}},\ \bibinfo {pages} {587}
  (\bibinfo {year} {1983})}\BibitemShut {NoStop}%
\bibitem [{\citenamefont {Schmid}(1983)}]{Schmid}%
  \BibitemOpen
  \bibfield  {author} {\bibinfo {author} {\bibfnamefont {A.}~\bibnamefont
  {Schmid}},\ }\href {\doibase 10.1103/PhysRevLett.51.1506} {\bibfield
  {journal} {\bibinfo  {journal} {Phys. Rev. Lett.}\ }\textbf {\bibinfo
  {volume} {51}},\ \bibinfo {pages} {1506} (\bibinfo {year}
  {1983})}\BibitemShut {NoStop}%
\bibitem [{\citenamefont {Chakravarty}(1982)}]{Chakravarty}%
  \BibitemOpen
  \bibfield  {author} {\bibinfo {author} {\bibfnamefont {S.}~\bibnamefont
  {Chakravarty}},\ }\href {\doibase 10.1103/PhysRevLett.49.681} {\bibfield
  {journal} {\bibinfo  {journal} {Phys. Rev. Lett.}\ }\textbf {\bibinfo
  {volume} {49}},\ \bibinfo {pages} {681} (\bibinfo {year} {1982})}\BibitemShut
  {NoStop}%
\bibitem [{\citenamefont {Fisher}\ and\ \citenamefont
  {Zwerger}(1985)}]{FisherZwerger}%
  \BibitemOpen
  \bibfield  {author} {\bibinfo {author} {\bibfnamefont {M.~P.~A.}\
  \bibnamefont {Fisher}}\ and\ \bibinfo {author} {\bibfnamefont
  {W.}~\bibnamefont {Zwerger}},\ }\href@noop {} {\bibfield  {journal} {\bibinfo
   {journal} {Phys. Rev. B}\ }\textbf {\bibinfo {volume} {32}},\ \bibinfo
  {pages} {6190} (\bibinfo {year} {1985})}\BibitemShut {NoStop}%
\bibitem [{\citenamefont {Leggett}\ \emph {et~al.}(1987)\citenamefont
  {Leggett}, \citenamefont {Chakravarty}, \citenamefont {Dorsey}, \citenamefont
  {Fisher}, \citenamefont {Garg},\ and\ \citenamefont
  {Zwerger}}]{LeggettReview}%
  \BibitemOpen
  \bibfield  {author} {\bibinfo {author} {\bibfnamefont {A.~J.}\ \bibnamefont
  {Leggett}}, \bibinfo {author} {\bibfnamefont {S.}~\bibnamefont
  {Chakravarty}}, \bibinfo {author} {\bibfnamefont {A.~T.}\ \bibnamefont
  {Dorsey}}, \bibinfo {author} {\bibfnamefont {M.~P.}\ \bibnamefont {Fisher}},
  \bibinfo {author} {\bibfnamefont {A.}~\bibnamefont {Garg}}, \ and\ \bibinfo
  {author} {\bibfnamefont {W.}~\bibnamefont {Zwerger}},\ }\href@noop {}
  {\bibfield  {journal} {\bibinfo  {journal} {Rev. Mod. Phys.}\ }\textbf
  {\bibinfo {volume} {59}},\ \bibinfo {pages} {1} (\bibinfo {year}
  {1987})}\BibitemShut {NoStop}%
\bibitem [{\citenamefont {Kane}\ and\ \citenamefont
  {Fisher}(1992)}]{KaneFisher}%
  \BibitemOpen
  \bibfield  {author} {\bibinfo {author} {\bibfnamefont {C.~L.}\ \bibnamefont
  {Kane}}\ and\ \bibinfo {author} {\bibfnamefont {M.~P.~A.}\ \bibnamefont
  {Fisher}},\ }\href@noop {} {\bibfield  {journal} {\bibinfo  {journal} {Phys.
  Rev. B}\ }\textbf {\bibinfo {volume} {46}},\ \bibinfo {pages} {15233}
  (\bibinfo {year} {1992})}\BibitemShut {NoStop}%
\bibitem [{\citenamefont {Torre}\ \emph {et~al.}(2010)\citenamefont {Torre},
  \citenamefont {Demler}, \citenamefont {Giamarchi},\ and\ \citenamefont
  {Altman}}]{us-nature}%
  \BibitemOpen
  \bibfield  {author} {\bibinfo {author} {\bibfnamefont {E.~D.}\ \bibnamefont
  {Torre}}, \bibinfo {author} {\bibfnamefont {E.}~\bibnamefont {Demler}},
  \bibinfo {author} {\bibfnamefont {T.}~\bibnamefont {Giamarchi}}, \ and\
  \bibinfo {author} {\bibfnamefont {E.}~\bibnamefont {Altman}},\ }\href@noop {}
  {\bibfield  {journal} {\bibinfo  {journal} {Nature Physics}\ }\textbf
  {\bibinfo {volume} {6}},\ \bibinfo {pages} {806} (\bibinfo {year}
  {2010})}\BibitemShut {NoStop}%
\bibitem [{\citenamefont {Dalla~Torre}\ \emph {et~al.}(2012)\citenamefont
  {Dalla~Torre}, \citenamefont {Demler}, \citenamefont {Giamarchi},\ and\
  \citenamefont {Altman}}]{us-prb}%
  \BibitemOpen
  \bibfield  {author} {\bibinfo {author} {\bibfnamefont {E.~G.}\ \bibnamefont
  {Dalla~Torre}}, \bibinfo {author} {\bibfnamefont {E.}~\bibnamefont {Demler}},
  \bibinfo {author} {\bibfnamefont {T.}~\bibnamefont {Giamarchi}}, \ and\
  \bibinfo {author} {\bibfnamefont {E.}~\bibnamefont {Altman}},\ }\href
  {\doibase 10.1103/PhysRevB.85.184302} {\bibfield  {journal} {\bibinfo
  {journal} {Phys. Rev. B}\ }\textbf {\bibinfo {volume} {85}},\ \bibinfo
  {pages} {184302} (\bibinfo {year} {2012})}\BibitemShut {NoStop}%
\bibitem [{\citenamefont {Mitra}\ and\ \citenamefont
  {Giamarchi}(2011)}]{Mitra-short}%
  \BibitemOpen
  \bibfield  {author} {\bibinfo {author} {\bibfnamefont {A.}~\bibnamefont
  {Mitra}}\ and\ \bibinfo {author} {\bibfnamefont {T.}~\bibnamefont
  {Giamarchi}},\ }\href {\doibase 10.1103/PhysRevLett.107.150602} {\bibfield
  {journal} {\bibinfo  {journal} {Phys. Rev. Lett.}\ }\textbf {\bibinfo
  {volume} {107}},\ \bibinfo {pages} {150602} (\bibinfo {year}
  {2011})}\BibitemShut {NoStop}%
\bibitem [{\citenamefont {Mitra}\ and\ \citenamefont
  {Giamarchi}(2012)}]{Mitra-long}%
  \BibitemOpen
  \bibfield  {author} {\bibinfo {author} {\bibfnamefont {A.}~\bibnamefont
  {Mitra}}\ and\ \bibinfo {author} {\bibfnamefont {T.}~\bibnamefont
  {Giamarchi}},\ }\href@noop {} {\bibfield  {journal} {\bibinfo  {journal}
  {Phys. Rev. B}\ }\textbf {\bibinfo {volume} {85}},\ \bibinfo {pages} {075117}
  (\bibinfo {year} {2012})}\BibitemShut {NoStop}%
\bibitem [{\citenamefont {Ingold}\ and\ \citenamefont
  {Nazarov}(1992)}]{IngoldNazarov}%
  \BibitemOpen
  \bibfield  {author} {\bibinfo {author} {\bibfnamefont {G.-L.}\ \bibnamefont
  {Ingold}}\ and\ \bibinfo {author} {\bibfnamefont {Y.~V.}\ \bibnamefont
  {Nazarov}},\ }\href
  {http://www.citebase.org/abstract?id=oai:arXiv.org:cond-mat/0508728}
  {\bibfield  {journal} {\bibinfo  {journal} {NATO ASI SERIES B}\ }\textbf
  {\bibinfo {volume} {294}},\ \bibinfo {pages} {21} (\bibinfo {year}
  {1992})}\BibitemShut {NoStop}%
\bibitem [{\citenamefont {{Devoret}}(1997)}]{Devoret}%
  \BibitemOpen
  \bibfield  {author} {\bibinfo {author} {\bibfnamefont {M.~H.}\ \bibnamefont
  {{Devoret}}},\ }in\ \href@noop {} {\emph {\bibinfo {booktitle} {Fluctuations
  Quantiques/Quantum Fluctuations}}},\ \bibinfo {editor} {edited by\ \bibinfo
  {editor} {\bibnamefont {{S.~Reynaud, E.~Giacobino, \& J.~Zinn-Justin}}}}\
  (\bibinfo {year} {1997})\ p.\ \bibinfo {pages} {351}\BibitemShut {NoStop}%
\bibitem [{\citenamefont {Barends}\ \emph {et~al.}(2011)\citenamefont
  {Barends}, \citenamefont {Wenner}, \citenamefont {Lenander}, \citenamefont
  {Chen}, \citenamefont {Bialczak}, \citenamefont {Kelly}, \citenamefont
  {Lucero}, \citenamefont {O'Malley}, \citenamefont {Mariantoni}, \citenamefont
  {Sank}, \citenamefont {Wang}, \citenamefont {White}, \citenamefont {Yin},
  \citenamefont {Zhao}, \citenamefont {Cleland}, \citenamefont {Martinis},\
  and\ \citenamefont {Baselmans}}]{Barends}%
  \BibitemOpen
  \bibfield  {author} {\bibinfo {author} {\bibfnamefont {R.}~\bibnamefont
  {Barends}}, \bibinfo {author} {\bibfnamefont {J.}~\bibnamefont {Wenner}},
  \bibinfo {author} {\bibfnamefont {M.}~\bibnamefont {Lenander}}, \bibinfo
  {author} {\bibfnamefont {Y.}~\bibnamefont {Chen}}, \bibinfo {author}
  {\bibfnamefont {R.~C.}\ \bibnamefont {Bialczak}}, \bibinfo {author}
  {\bibfnamefont {J.}~\bibnamefont {Kelly}}, \bibinfo {author} {\bibfnamefont
  {E.}~\bibnamefont {Lucero}}, \bibinfo {author} {\bibfnamefont
  {P.}~\bibnamefont {O'Malley}}, \bibinfo {author} {\bibfnamefont
  {M.}~\bibnamefont {Mariantoni}}, \bibinfo {author} {\bibfnamefont
  {D.}~\bibnamefont {Sank}}, \bibinfo {author} {\bibfnamefont {H.}~\bibnamefont
  {Wang}}, \bibinfo {author} {\bibfnamefont {T.~C.}\ \bibnamefont {White}},
  \bibinfo {author} {\bibfnamefont {Y.}~\bibnamefont {Yin}}, \bibinfo {author}
  {\bibfnamefont {J.}~\bibnamefont {Zhao}}, \bibinfo {author} {\bibfnamefont
  {A.~N.}\ \bibnamefont {Cleland}}, \bibinfo {author} {\bibfnamefont {J.~M.}\
  \bibnamefont {Martinis}}, \ and\ \bibinfo {author} {\bibfnamefont {J.~J.~A.}\
  \bibnamefont {Baselmans}},\ }\href {\doibase 10.1063/1.3638063} {\bibfield
  {journal} {\bibinfo  {journal} {Applied Physics Letters}\ }\textbf {\bibinfo
  {volume} {99}},\ \bibinfo {eid} {113507} (\bibinfo {year}
  {2011})}\BibitemShut {NoStop}%
\bibitem [{\citenamefont {Haldane}(1981)}]{Haldane}%
  \BibitemOpen
  \bibfield  {author} {\bibinfo {author} {\bibfnamefont {F.~D.~M.}\
  \bibnamefont {Haldane}},\ }\href {\doibase 10.1103/PhysRevLett.47.1840}
  {\bibfield  {journal} {\bibinfo  {journal} {Phys. Rev. Lett.}\ }\textbf
  {\bibinfo {volume} {47}},\ \bibinfo {pages} {1840} (\bibinfo {year}
  {1981})}\BibitemShut {NoStop}%
\bibitem [{\citenamefont {Giamarchi}(2004)}]{Giamarchi}%
  \BibitemOpen
  \bibfield  {author} {\bibinfo {author} {\bibfnamefont {T.}~\bibnamefont
  {Giamarchi}},\ }\href@noop {} {\emph {\bibinfo {title} {Quantum Physics in
  One Dimension}}}\ (\bibinfo  {publisher} {Oxford University Press},\ \bibinfo
  {address} {Oxford},\ \bibinfo {year} {2004})\BibitemShut {NoStop}%
\bibitem [{\citenamefont {Betz}\ \emph {et~al.}(2011)\citenamefont {Betz},
  \citenamefont {Manz}, \citenamefont {B\"ucker}, \citenamefont {Berrada},
  \citenamefont {Koller}, \citenamefont {Kazakov}, \citenamefont {Mazets},
  \citenamefont {Stimming}, \citenamefont {Perrin}, \citenamefont {Schumm},\
  and\ \citenamefont {Schmiedmayer}}]{Schmiedmayer2}%
  \BibitemOpen
  \bibfield  {author} {\bibinfo {author} {\bibfnamefont {T.}~\bibnamefont
  {Betz}}, \bibinfo {author} {\bibfnamefont {S.}~\bibnamefont {Manz}}, \bibinfo
  {author} {\bibfnamefont {R.}~\bibnamefont {B\"ucker}}, \bibinfo {author}
  {\bibfnamefont {T.}~\bibnamefont {Berrada}}, \bibinfo {author} {\bibfnamefont
  {C.}~\bibnamefont {Koller}}, \bibinfo {author} {\bibfnamefont
  {G.}~\bibnamefont {Kazakov}}, \bibinfo {author} {\bibfnamefont {I.~E.}\
  \bibnamefont {Mazets}}, \bibinfo {author} {\bibfnamefont {H.-P.}\
  \bibnamefont {Stimming}}, \bibinfo {author} {\bibfnamefont {A.}~\bibnamefont
  {Perrin}}, \bibinfo {author} {\bibfnamefont {T.}~\bibnamefont {Schumm}}, \
  and\ \bibinfo {author} {\bibfnamefont {J.}~\bibnamefont {Schmiedmayer}},\
  }\href {\doibase 10.1103/PhysRevLett.106.020407} {\bibfield  {journal}
  {\bibinfo  {journal} {Phys. Rev. Lett.}\ }\textbf {\bibinfo {volume} {106}},\
  \bibinfo {pages} {020407} (\bibinfo {year} {2011})}\BibitemShut {NoStop}%
\end{thebibliography}


\end{document}